\documentclass[%
 reprint,
 amsmath,amssymb,
 aps
floatfix,
]{revtex4-1}
\usepackage{graphicx}% Include figure files
\usepackage{dcolumn}% Align table columns on decimal point
\usepackage{bm}% bold math
\usepackage{color}
\usepackage[utf8]{inputenc}
\usepackage{capt-of}
\usepackage{tabu}
\usepackage{mwe} % For dummy images
\usepackage{subcaption}
\usepackage{soul}
\usepackage{xr}
\makeatletter
\newcommand*{\addFileDependency}[1]{% argument=file name and extension
  \typeout{(#1)}
  \@addtofilelist{#1}
  \IfFileExists{#1}{}{\typeout{No file #1.}}
}
\makeatother

\begin{document}

%PREVIOUS:
%\title{Self-organized clustering of K-Nearest-Neighbor self-propelled particle systems that impedes global polarization}

%CH's SUGGESTED:
%\title{Emergent Metric-like Collective States of Clusters of Active Particles\\
%with Metric-free Polar Interactions}
% PR suggested:
%\title{Clustering \& Emergent Metric-like Collective States of Active Particles\\
%with Metric-free Polar Alignment}
% CH new suggestion (Aug 8): 
\title{Emergent Metric-like States of Active Particles with Metric-free Polar Alignment}

% MUST INCLUDE: Clustering | Metric-free/non-metric alignment or so | active | (discontinuous phase transition) | (metric particle)

% Clustering induces a discontinuous phase transition in an active system with a metric-free alignment
% Clustering determines an active system with a metric-free alignment

% OTHER POTENTIAL TITLES:
%___ Self-organized clustering of K-Nearest-Neighbor active systems alters the nature of phase transition
%___ Self-organized clustering of K-Nearest-Neighbor active system induces a discontinuous phase transition
%___ Self-organization of K-Nearest-Neighbor active system induces a discontinuous phase transition
%___ Self-organized clustering of non-metric active system induces a discontinuous phase transition
%___ Self-organized clustering of an active system with a non-metric interaction network induces a discontinuous phase transition
%___ Self-organized clustering of an active system with a K-Nearest-Neighbor interaction network induces a discontinuous phase transition

\author{Yinong Zhao$^{1,2}$, Cristi\'an Huepe$^{4,5,6}$, and Pawel Romanczuk$^{1,2,3}$}

\address{$^1$Institute for Theoretical Biology, Department of Biology, Humboldt-Universität zu Berlin, 10115 Berlin, Germany}
\address{$^2$Bernstein Center for Computational Neuroscience Berlin, 10115 Berlin, Germany}
\address{$^3$Science of Intelligence, Research Cluster of Excellence, Marchstr. 23, 10587 Berlin, https://www.scienceofintelligence.de}
\address{$^4$School of Systems Science, Beijing Normal University, Beijing 100875, China}
\address{$^5$CHuepe Labs, 2713 W Haddon Ave \#1, Chicago, IL 60622, USA}
\address{$^6$Northwestern Institute on Complex Systems and ESAM, Northwestern University, Evanston, IL 60208, USA}

\begin{abstract}
We study a model of self-propelled particles interacting with their $k$ nearest neighbors through polar alignment.
By exploring its phase space as a function of two nondimensional parameters (alignment strength $g$ and Peclet number $\mathrm{Pe}$), we identify two distinct order-disorder transitions.
One is continuous, occurs at a low critical $g$ value independent of Pe, and resembles a mean-field transition with no density-order coupling.
The other is discontinuous, depends on a combined control parameter involving $g$ and Pe, and results from the formation of small, dense, highly persistent clusters of particles that follow metric-like dynamics. 
These dense clusters form at a critical value of the combined control parameter $\mathrm{Pe}/g^{\alpha}$, with $\alpha \approx 1.5$, which appears to be valid for different alignment-based models.
%
%We observe a clustering process that is determined solely by $\mathrm{Pe}/g^{\alpha}$ where $\alpha\approx1.5$ appears to be an exponent generically valid for the model and its variants.
%
Our study shows that models of active particles with metric-free interactions can produce characteristic length-scales and self-organize into metric-like collective states that undergo metric-like transitions.
%
%Our study reveal a generic feather of SPP model based on k-nearest-neighbor interaction network, we close the gap between previous studies and provide a new "cluster" perspective to understand collective behavior.
\end{abstract}

\maketitle

%% Introduction
% \section{Introduction}

%Self-driven living matter systems are ubiquitously observed in nature \cite{theraulaz2002spatial, bazazi2011nutritional,DeThViReview2012,cavagna2010scale,cavagna2015short,ballerini2008interaction,Breder_54, couzin05}. Based on short-range interaction between individuals, a group is able to perform global order. One of the most prominent model for describing the collective motion phenomena is the Vicsek model \cite{vicsek1995novel}, in which local orientational alignment between self-propelled particles leads to a global orientational order.
%
%Self-organization can be ubiquitously observed in living systems cite{theraulaz2002spatial, bazazi2011nutritional,DeThViReview2012,cavagna2010scale,cavagna2015short,ballerini2008interaction,Breder_54, couzin05}. 
Over the past few decades, great progress has been made in understanding one of the simplest forms of biological self-organization: the emergence of large-scale collective movement due to local interactions between active agents \cite{theraulaz2002spatial, bazazi2011nutritional,DeThViReview2012,cavagna2010scale,cavagna2015short,ballerini2008interaction,Breder_54, couzin05}.
A prominent model for describing this phenomenon is the Vicsek model \cite{vicsek1995novel}, in which self-propelled particles achieve global polar order through local alignment interactions between neighbors within a given radius. 
In general, Vicsek-like models typically consider {\emph{metric}} interactions, where alignment forces depend on the Euclidean distance between particles \cite{chate2008collective,bertin2006boltzmann,ginelli2016physics}.
This leads to a coupling between local density and local order that has been shown to result in various features of the ordered phase, such as the presence of anomalous density fluctuations or the emergence of moving density bands produced by long wavelength instabilities near the order-disorder transition \cite{bertin2006boltzmann,ginelli2016physics,zhao2021phases}.

An alternative form of Vicsek-like models, inspired by empirical observations, defines interacting agents not in terms of their metric distance, but instead as a function of their {\emph{topological}} distance in an interaction network that does not depend on geometrical proximity \cite{ballerini2008interaction,ginelli2010relevance}.
In the so-called KNN models, for example, each agent interacts with its $k$ nearest neighbors (KNN), regardless of their actual Euclidean distance \cite{ballerini2008interaction,chou2012kinetic,rahmani2021topological,martin2021fluctuation}.
It has been commonly believed that, since such topological interactions contain no direct density-order coupling, they should not induce any density instability, producing instead a homogeneous flocking phase at the onset of order \cite{ginelli2010relevance,peshkov2012continuous,chou2012kinetic}.
However, recent results have shown that moving density bands can in fact appear in a KNN Vicsek-like model, and that these can be amplified in spatially heterogeneous environments \cite{martin2021fluctuation,rahmani2021topological}.
This has led to the proposal of different mechanisms that could produce an effective coupling between density and order in metric-free flocking models \cite{martin2021fluctuation,rahmani2021topological}.
Despite these efforts, no clear picture of how such spatial structures can appear in this type of systems has emerged.
%that there are still many open questions regarding the self-organization of active systems with topological interactions. 
%

In this Letter, we investigate the order-disorder transitions in a KNN Vicsek-like model.
%in which each self-propelled particle tends to align with its $k$ nearest neighbors irrespective of their Euclidean distance.
%
By carrying out a systematic analysis of its collective states as a function of two dimensionless parameters (the effective coupling strength $g$ and Peclet number Pe), we reveal the presence of two distinct order-disorder transitions, one for low $g$ and one for high $g$.
We show that the former resembles a mean-field transition with no density-order coupling, whereas the latter is related to the formation of dense persistent clusters that behave at a coarse-grained level like self-propelled ``particles'' with metric alignment interactions, which then develop a density-order coupling at large spatiotemporal scales.
%
%Our results provide a better understanding of the collective states that can emerge in models with topological interactions and their relationship to those found in metric systems.
Our results provide important insights on the emergent collective states due to topological interactions and their relationship to those found in metric systems.

% \section{Model description}
% \label{sec:model3}
% Results
% \section{Simulation}

We consider $N$ self-propelled particles moving continuously in a 2-dimensional arena of size $L\times L$, with periodic boundary conditions.
At time $t$, particle $i$ is located at position $\vec{r}_i(t)$ and advances with constant speed $v_0$, following
\begin{equation}
    \label{eq:knn-model-location}
    \frac{d\vec{r}_i(t)}{dt} = v_0\hat{n}_i(t),
\end{equation}
where $\hat{n}_i(t) = \left[\cos\theta_i(t),\sin\theta_i(t) \right]^T$ is a unit vector pointing in the heading direction of particle $i$.
The angle $\theta_i(t)$ tends to align to its neighbors through the interaction dynamics defined by
\begin{equation}
    \label{eq:knn-model-angle}
    \frac{d\theta_i(t)}{dt} = \frac{1}{\tau}\left\langle \mathrm{mod}^{*}(\theta_j -\theta_i) \right\rangle_{j\in S_i} + \sigma\xi_{\theta}.
\end{equation}
Here, $S_i$ is a set of cardinality $k+1$ that contains $i$ itself and the indexes of all $k$ nearest neighbors of particle $i$, 
the alignment relaxation time $\tau$ is inversely proportional to the strength of the alignment interaction, 
the modified modulo function $\mathrm{mod}^{*}(x)=\mathrm{mod}(x+\pi,2\pi)-\pi$ is defined in terms of the standard modulo function $\mathrm{mod}$,
the random variable $\xi_{\theta}$ introduces uncorrelated Gaussian white noise, and $\sigma$ determines the noise strength. 
We characterize the parameter space in terms of two nondimensional quantities \cite{martin2018collective,zhao2021phases}: 
the dimensionless alignment strength 
$g = 1/(\tau\sigma^2)$ and the Peclet number 
$\mathrm{Pe} = v_0\sqrt{\rho}/\sigma^2$,
where $\rho = N/L^2$ is the mean density.
In practice, we explore the collective phases by computing the statistically stationary states obtained for different values of the angular noise $\sigma$ and alignment relaxation time $\tau$, plotting the standard polarization order parameter $\Phi=|\sum_{i = 1}^{N}\hat{n}(\theta_i)|/N$ to evaluate the degree of alignment order for each parameter combination.

% \section{Results}
% \label{sec:results3}

\begin{figure}[t]
    \includegraphics[width=0.48\columnwidth]{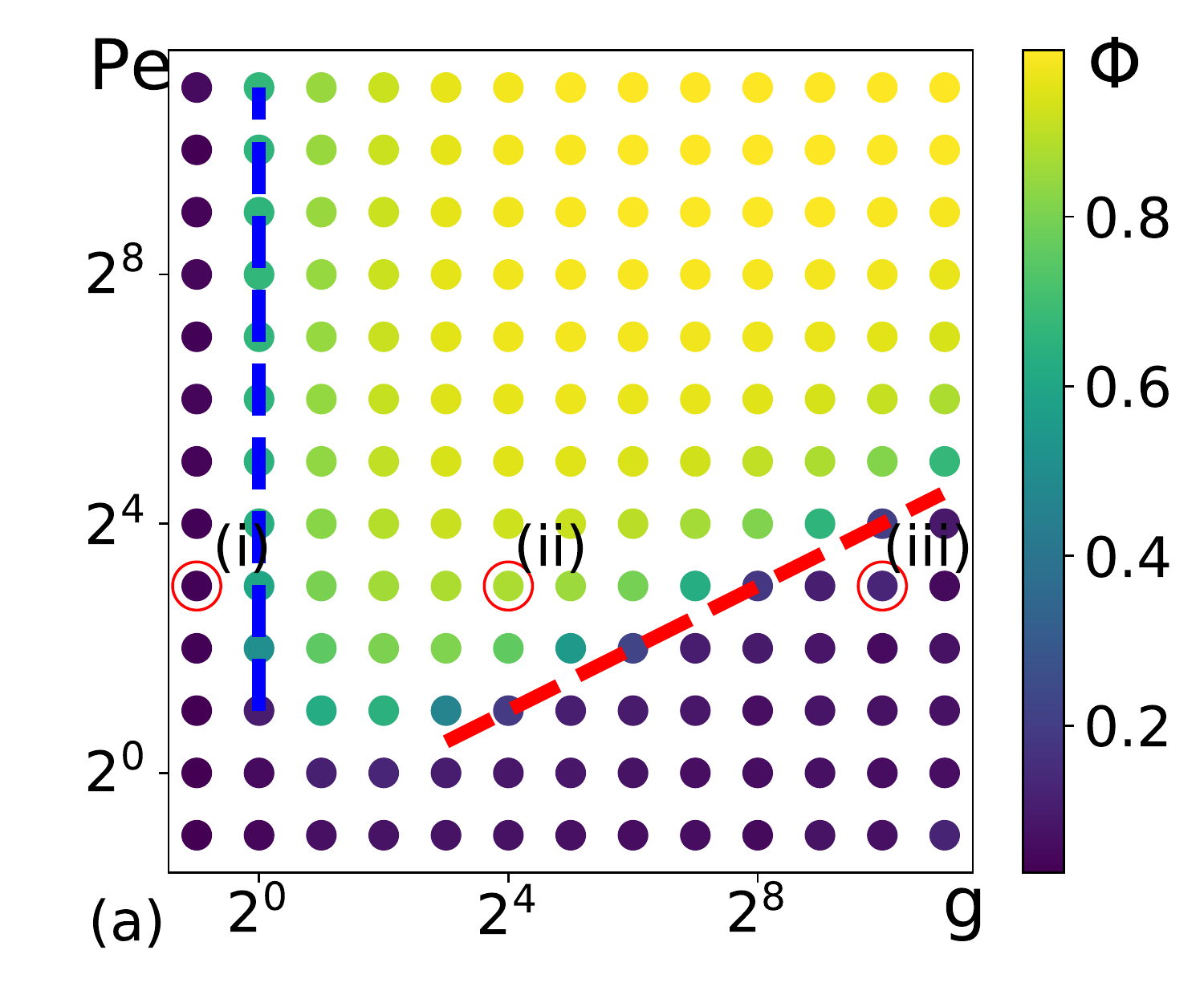}
    %   \caption{}
      \includegraphics[width=0.48\columnwidth]{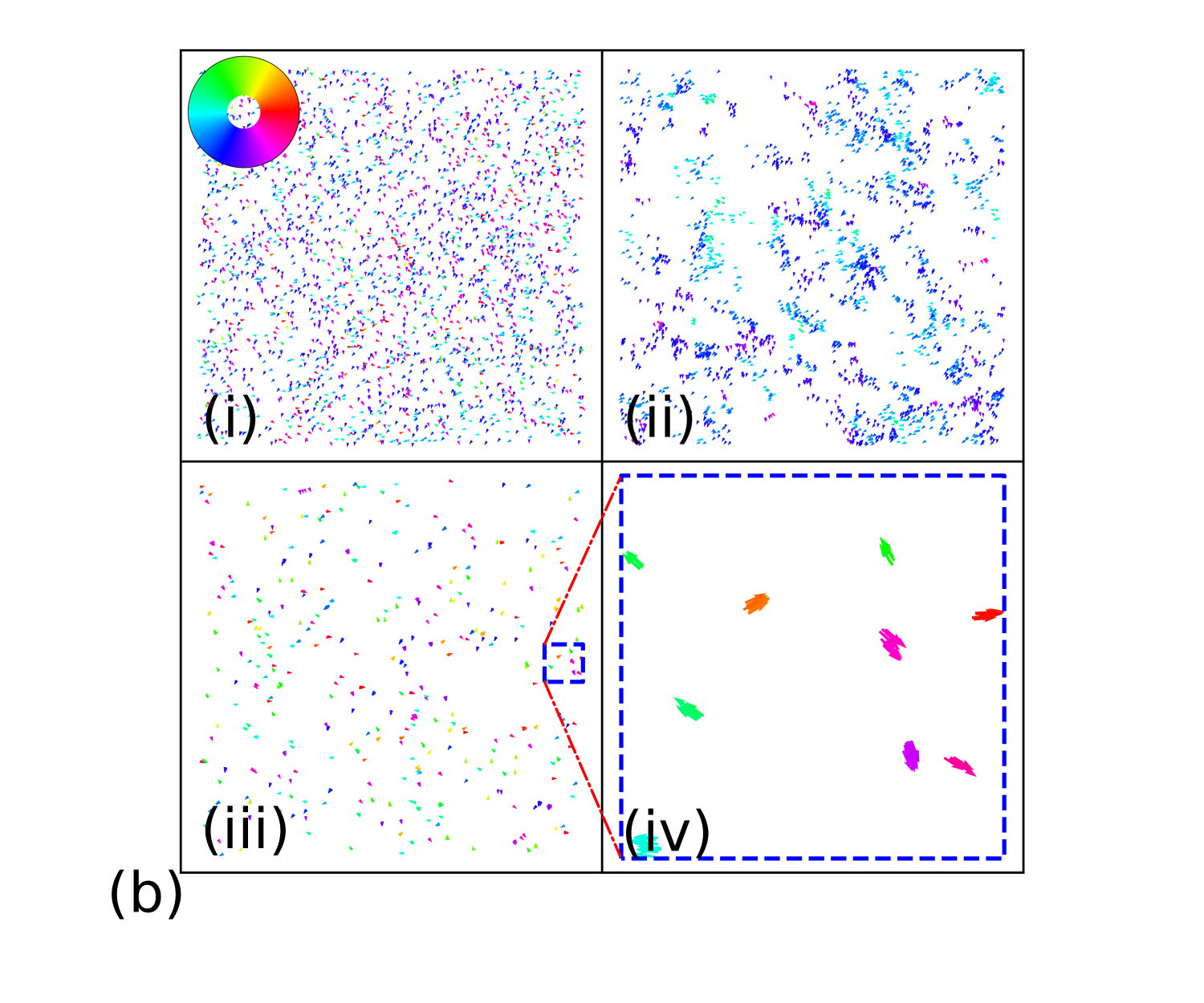}
      \includegraphics[width=0.48\columnwidth]{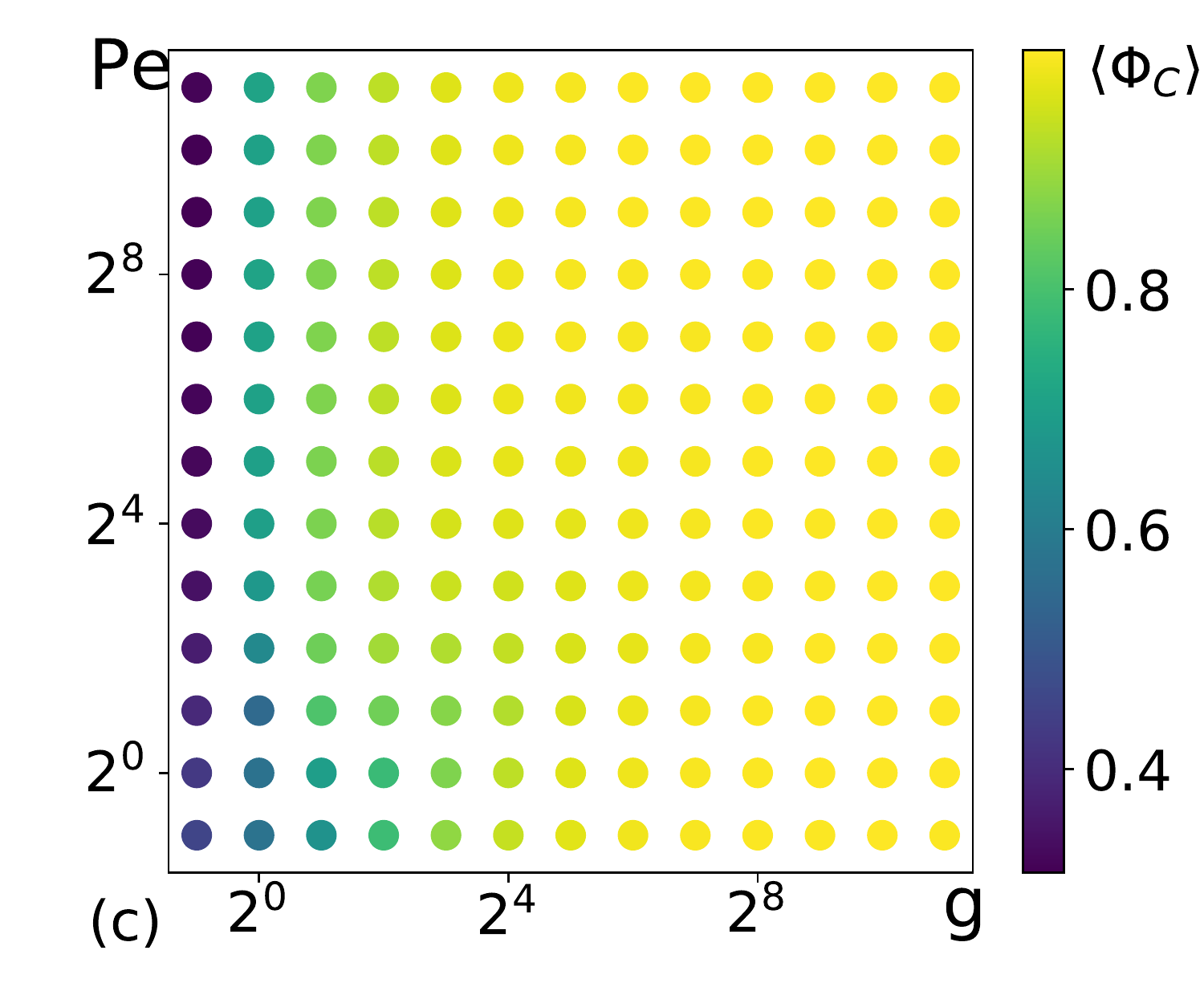}
    %   \caption{}
      \includegraphics[width=0.48\columnwidth]{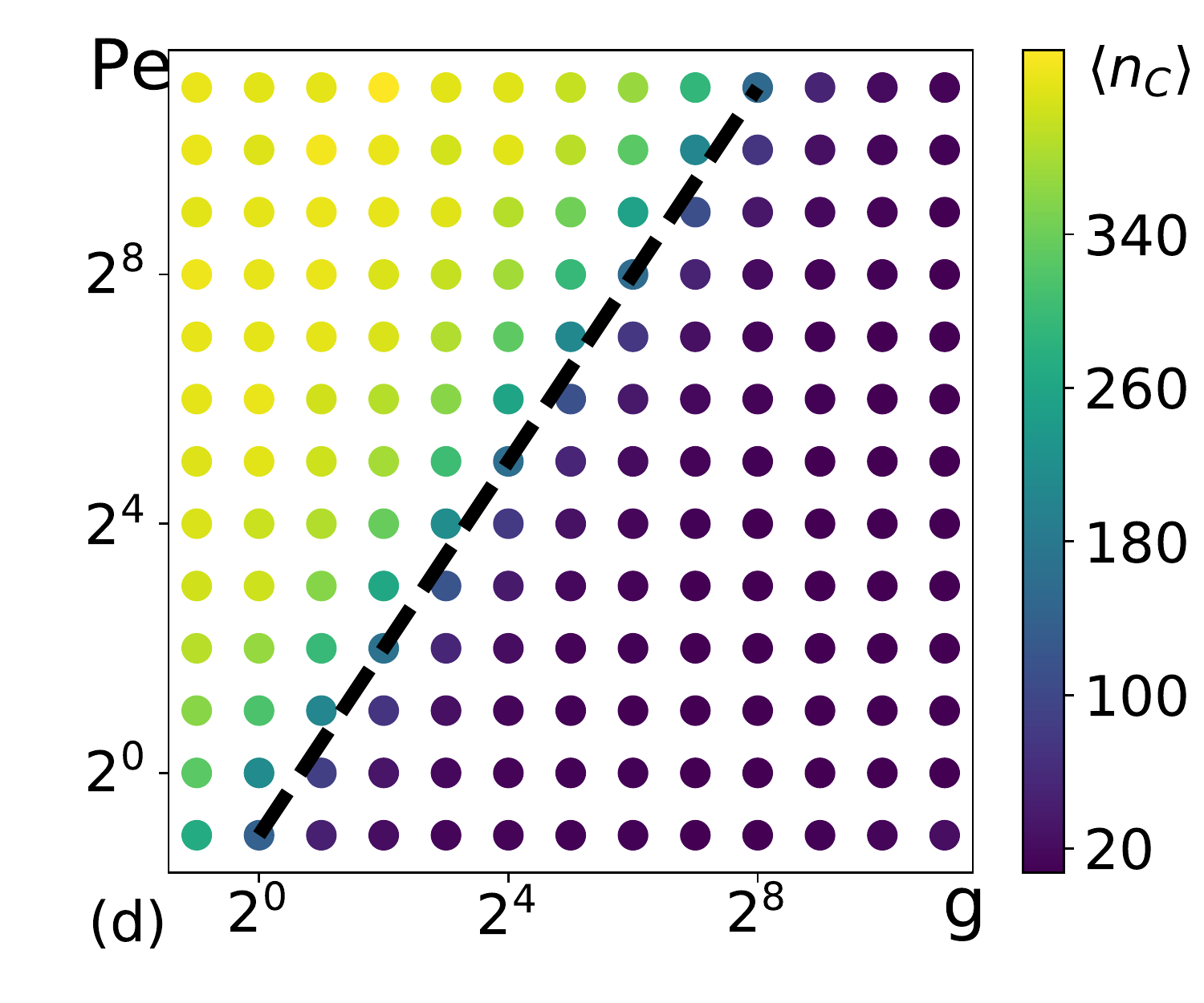}
    \caption[Diagrams]{Phase diagrams as a function of ($g$,Pe) and related snapshots, for $k=3$, $N=2500$, $\rho=N/L^2=1$ and $v_0=0.2$. All order parameters are averaged over 1000 frames after reaching the stationary state. (a) Mean global polar order, with the blue and red dashed lines showing the order-disorder transitions A and B, respectively. (b) Snapshots of the three states labeled in panel a, where arrows represent particles and are colored by heading angle according to the inserted color disk, for $\mathrm{Pe}=8$ and (i) $g=0.5$, (ii) $g=16$, (iii) $g=1024$, and (iv) is a zoom-in of the blue dashed square in iii. (c) Mean polar order of clusters. (d) Mean cluster size with a black dashed line showing the transition from large ($\langle n_C \rangle > 125$) to small ($\langle n_C \rangle < 125$) clusters.}
    \label{fig:diagrams}
\end{figure}

Figure \ref{fig:diagrams}(a) presents the order-disorder phase diagram as a function of $g$ and Pe, for a set of simulations with $N=2500$, $v_0=0.2$, $L=50$, $k=3$, and $\rho=1$. 
The color scale displays the mean $\Phi$, averaged over time after a stationary state is reached.
This diagram shows that the system can undergo two different transitions as a function of $g$, for constant Pe.
The first one, which we will refer to as transition A (sketched by a blue dashed line), appears in the weak alignment regime ($g \approx 1$).
It occurs when alignment forces become strong enough to overcome noise, which results in a transition from a disordered to an aligned state.
The second one, transition B (sketched as a red dashed line), reflects the reemergence of a disordered state as $g$ is increased to high values.
Its origin is thus more counter-intuitive, since it implies that increasing the inter-particle alignment strength can result in a loss of polar order.
Both transitions appear as straight lines in the log-log phase diagram. Indeed, analytical calculations \cite{chou2012kinetic} show that transition A occurs at fixed $g$, for any Pe value, and our numerical analysis shows that transition B also forms a line, corresponding to a power-law relationship between $g$ and Pe \cite{SI}.
%It appears that the both transition boundaries are straight lines in the log-log phase diagram.

An examination of both transitions shows that they are of a fundamentally different nature.
Transition A corresponds to a non-equilibrium, continuous symmetry-breaking transition controlled only by $g$, which is well described by a mean-field kinetic theory \cite{chou2012kinetic}. 
In large scale simulations, we verified that this transition appears as continuous in systems of up to $N=10^6$ particles.
By contrast, transition B appears to be discontinuous, showing a bimodal distribution of instantaneous values of the polarization order parameter $\Phi$ in the stationary state. 
We also observe large-scale density bands near its critical point \cite{SI,gregoire2004onset}, which have been linked to the discontinuous nature of the order-disorder transition in metric Vicsek-like models \cite{chate2008collective,caussin2014emergent,solon2015phase}.

%-------------------------------------------------------------- FIGURE 1
%-------------------------------------------------------------- FIGURE 1

%In Fig.~\ref{fig:diagrams}b we show snapshots of stationary states for constant Peclet number, $\mathrm{Pe}=8$, but different coupling strengths $g$ as marked in the Panel a: (i) $g=0.5$, (ii) $g=16$, (iii) $g=1024$. Panel b(iv) is a zoom-in of Panel b(iii), showing that the particles form clusters that are polarized internally but not aligned to other clusters in the high g low Pe regime. 
%It leads us to study the clusters. We define clusters in the term of weakly connected graph: the interaction network is a directed graph where particle $j$ belonging to the k-nearest neighborhood of particle $i$ is represented by an arrow from vertex $j$ to vertex $i$; vertex $i$ and $j$ is considered connected if there exists an arrow between them, and a subgraph where all vertices are connected to each other by some path corresponds a cluster. Specifically, particle $j$ is considered to belong the cluster K if there exists any particle $i$ in cluster K satisfying that particle $j$ is included in particle $i$'s $k$ nearest neighbors or particle $i$ is a member of particle $j$'s k nearest neighbors. 

In order to explore the origin of transition B, we analyzed the spatial distribution of particles as we increase the coupling strength $g$, for a fixed Peclet Number Pe.
Figure \ref{fig:diagrams}(b) presents snapshots of three stationary states of the simulations used to compute the points labeled (i), (ii), and (iii) in the phase diagram in panel (a). 
For low $g$ (i), individual particles head in different directions and the system is disordered. 
For intermediate $g$ (ii), particles align and the system becomes ordered while diffuse clusters are formed.
For larger $g$ values (iii), strong alignment interactions result in the formation of dense, persistent clusters that are composed of highly aligned particles but are not aligned to each other.
Transition B therefore appears to result from the formation of clusters that become more dense internally and more sparse with respect to each other, which leads them to rarely interact and to therefore lose global order.
We will analyze the relationship between these clusters and transition B in more detail below.

We begin by defining particle clusters in terms of an interaction network that connects node $i$ to node $j$ at a given time if particle $j$ is among the k-nearest neighbors of particle $i$.
Each cluster thus corresponds to a weakly connected component of the resulting directed network \cite{Newman2010}, that is, to a sub-graph in which all nodes are connected by a topological path that ignores the direction of the links.
We can then define $n_C$ and $\Phi_C$ as the number and polarization of the particles within each cluster.

Figures \ref{fig:diagrams}(c) and \ref{fig:diagrams}(d) respectively display the mean cluster polarization $\langle \Phi_C \rangle$ and mean cluster size $\langle n_{C} \rangle$ as a function of $g$ and Pe, for the same simulations presented in panel (a).
As the alignment strength $g$ is increased, we find that $\langle \Phi_C \rangle$ grows and $\langle n_{C} \rangle$ decreases.
For high enough $g$, the system thus always reaches a state where it forms small and persistent \emph{dense clusters} made of highly aligned particles.
We label this transition by a black dashed line that denotes the formation of clusters of mean size $\langle n_C \rangle < 125$, which we arbitrarily set as the small cluster threshold.
To the right of this line, each cluster becomes so dense and coherent that it can be viewed as acting as a collective unit.
Global polar order is then lost as $g$ is further increased and these units start heading in different directions, even though the particles within each cluster remain highly aligned.

%%%%%%%%%%%%%% Pawel until here %%%%%

\begin{figure}[t]
    \centering
    \includegraphics[width=0.95\columnwidth]{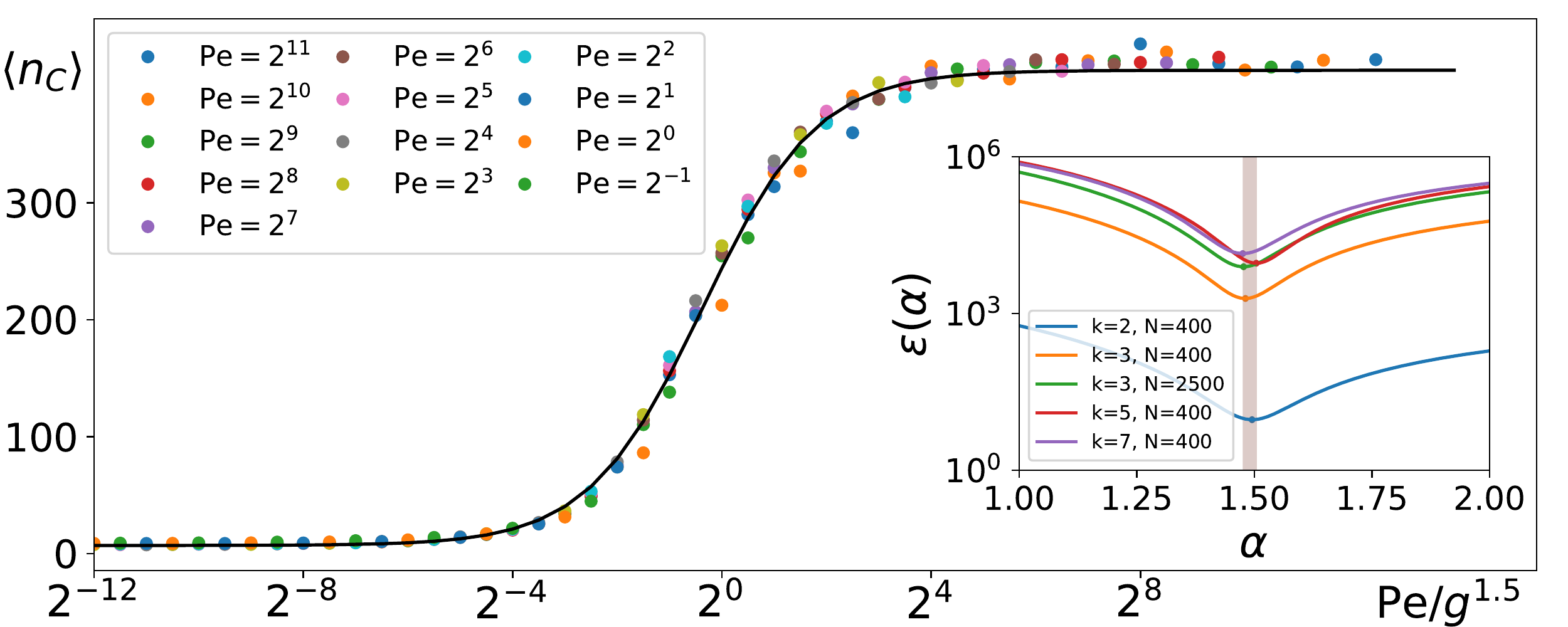}
    \caption[Mean cluster size vs.~$\mathrm{Pe}/g^{1.5}$]{
    Mean cluster size $\langle n_C \rangle$ vs. combined control parameter $\mathrm{Pe}/g^{\alpha}$ (with $\alpha=1.5$) for $k=3$ and $N=2500$.
    Thirteen sets of dots of different colors display the same $\langle n_C \rangle$ results presented in Fig.~\ref{fig:diagrams}d for different Pe values (listed in the top-left inset).
    With this parametrization, all dots collapse into a universal curve that characterizes the transition between large and small clusters, which we represent by an overlaid black smoothed curve (generated with the LOWESS method) that minimizes its mean square distance $\epsilon$ to all dots. The bottom-right inset shows the corresponding $\epsilon$ as a function of $\alpha$ for the $k$ and $N$ values listed in the inset. Their minima, marked by small dots, provides the $\alpha$ that best collapse the data (all found within the highlighted $1.49\pm0.015$ range).
    }
    \label{fig:nC-transition-1}
\end{figure}

Figure \ref{fig:nC-transition-1} shows that the mean cluster size $\langle n_C \rangle$ depends on a single combined control parameter, given by $\mathrm{Pe}/g^{\alpha}$ with $\alpha=1.5$, since all curves appear to collapse when we express the same results presented in Fig.\ref{fig:diagrams}(d) as a function of this quantity, without depending on the specific values of $\mathrm{Pe}$ or $g$.
To compute $\alpha$ more precisely, we search for the value that minimizes the mean square distance $\epsilon(\alpha)$ between all $\langle n_C \rangle$ values and their smoothed curve (generated through a LOWESS method \cite{LOWESS}) along the $\mathrm{Pe}/g^{\alpha}$ axis. 
% from where we find that $\alpha \approx 1.49\pm 0.015$.
%
% Furthermore, 
The inset in Fig.~\ref{fig:nC-transition-1} shows that, for different values of $k$ and $N$, all curves collapse when expressed in terms of this combined control parameter, with an $\alpha$ that minimizes $\epsilon(\alpha)$ and is always 
% appears to be close to $1.5$.
in the $\alpha \approx 1.49\pm 0.015$ range. 
For simplicity, we will therefore use $\alpha=1.5$ \cite{endnote1} in all our analyses below. 
We point out that this collapse of the $\langle n_C \rangle$ curves seems to hold for different systems sizes and alignment interaction functions.
In addition, we also find that the $\Phi$ curves for the global polar order also seem to collapse near transition B when express them as a function of $\mathrm{Pe}/g^{\beta}$, for $\beta \approx 0.45$ \cite{SI}. 

% defined as:
% \begin{equation}
%     \label{eq:sigmoid-type-fitting}
%     \langle n_C \rangle = \frac{\omega_1}{1+ \mathrm{e}^{-\omega_2(\log_2(\mathrm{Pe}/g^{\alpha})-\omega_3)}} + \omega_4
% \end{equation}
% % where $\omega_i$s denote factors to transform the sigmoid function to fit the ($\mathrm{Pe}/g^{\alpha}$,$\langle n_C \rangle$) points. And we find that $\alpha\approx 1.49\pm 0.01$ makes the points best fitted by the Eq.~\ref{eq:sigmoid-type-fitting}, qualified by the mean of squared residuals. 
% where $\omega_1\approx212.3$, $\omega_2\approx 1.0$, $\omega_3\approx 1.0$, $\omega_4\approx 8.4$ denote factors of the sigmoid-type function,

%-------------------------------------------------------------- FIGURE 2
%-------------------------------------------------------------- FIGURE 2

As described above, each dense cluster appears to effectively follow the dynamics of a single, highly persistent, self-propelled particle placed at its center of mass. 
Our simulations show that the shape of these clusters is statistically isotropic and that both their mean spatial extension and their mean size $\langle n_C \rangle$ are functions of $\mathrm{Pe}/g^{1.5}$.
We also find that the effective angular noise of a dense cluster is well approximated by $\tilde \sigma=\sigma/\sqrt{n_C}$, which is simply the combination of the independent random angular fluctuations of its constituents.
In order to compare the interactions between dense clusters to the interactions between individual particles, we systematically analyzed the binary scattering in simulations of collision between two clusters and between two particles, following the approach in \cite{hanke2013understanding}.
Although the detailed interaction between clusters can be complex, especially in collisions between different size clusters, we found that the mean effective interaction can be roughly described as an alignment force that only acts when the clusters touch and depends on their level of overlap.
Therefore, when representing clusters by effective particles, their interactions become short-ranged metric alignment forces that depend on the cluster sizes and distances \cite{SI}.

We will now describe transition B in terms of effective particles with metric interactions that represent the dense clusters, by defining a rescaled, coarse-grained ``cluster-based'' (CB) model.
To simplify our analyses, we will use identical effective particles, despite the fact that clusters display heterogeneous sizes and features.
In the CB model, we thus consider $\tilde{N}$ identical effective particles that follow the dynamics given by
\begin{eqnarray}
    \frac{d\vec{r}_i(t)}{dt} &=& \tilde{v}_0\hat{n}_i(t) + \vec{F}_i\label{eq:coarse-grained1}\\
    \frac{d\theta_i(t)}{dt} &=& \frac{1}{\tilde{\tau}}\left\langle \mathrm{mod}^{*}(\theta_j -\theta_i) \right\rangle_{j\in S_i} + \tilde{\sigma}\xi_{\theta},
    \label{eq:model-metric}    
\end{eqnarray}
where $\vec{r}_i(t)$ and $\theta_i(t)$ now represent the positions and orientations of the effective particles. 
The set of interacting neighbors is given here by 
$S_i=\left\{j \;|\; |\vec{r}_i-\vec{r}_j|\leq R \right\}$,
where $R$ is the effective interaction range. 
Parameters $\tilde{v_0}$, $\tilde{\tau}$, and $\tilde{\sigma}$ govern the individual effective particle dynamics, as in the original KNN model.
We are interested in avoiding in the CB model the further clustering that is known to occur in metric systems at low noise \cite{barberis2018emergence}, since each effective particle already represents a typical cluster in the KNN system. 
We thus introduce an additional phenomenological repulsive interaction $\vec{F}_i$, given by
\begin{equation}
\label{eq:repulsion-2}
    \vec{F}_i = \sum_{j \in S_i} \; 
		\frac{ \left\| \vec{r}_{ij} \right\| - R}{R} \frac{\vec{r}_{ij}}{\left\|\vec{r}_{ij}\right\|}.
\end{equation}
This corresponds to a linearly decaying force that vanishes at 
$\left\| \vec{r}_{ij} \right\| = R$. 
We verified that our results below do not significantly depend on our choice of $\vec{F}_i$, as long as it is strong enough to avoid clustering.

%This repulsive force reduces the interaction duration, thus leading to a scattering-like alignment dynamics as observed for clusters in the full KNN model.
%In the following, we will investigate in how far the coarse-grained ``cluster-based model" (CB model) is able to reproduce the large-scale behavior of the original KNN model.

We will now determine the parameters required for the effective particles in the CB model to reproduce the features of an average cluster in the KNN model. 
We begin by setting the total number of effective particles in the CB model, their mean speed, and their angular noise to $\tilde{N} = N/\langle n_C \rangle$,
$\tilde{v}_0\approx v_0$, and $\tilde{\sigma} = \sigma/\sqrt{\langle n_C\rangle}$, respectively. 
Each effective particle thus represents a typical dense cluster in the KNN model; it is composed of highly aligned particles and has a mean heading angle that fluctuates according to the combined, uncorrelated angular noise of all its constituent particles.
The remaining parameters, $R$ and $\tilde \tau $, cannot be directly estimated because they result from a variety of complex collisions between heterogeneous KNN clusters.
Instead, we can identify the $R$ and $\tilde{\tau}$ that best fit the corresponding KNN results.
To do this, we first make the ansatz that the effective interaction range is proportional to the mean cluster radius $\langle r_C \rangle$, which we expect to be a function of $\langle n_C \rangle$, and thus of $\mathrm{Pe}/g^{1.5}$.
We then assume a simple relation between $\tilde{\tau}$ and $\tau$, and between $R$ and $\langle r_C \rangle$ \cite{endnote2}, of the form
\begin{eqnarray}
    R &=& a \, \langle r_C\rangle \label{eq:R}\\
    \tilde{\tau} &=& f\left(\langle n_C \rangle, k\right)\,\tau,
    \label{eq:tau}
\end{eqnarray}
where $a$ is an unknown constant and $f$ is an unknown function of $\langle n_C \rangle$ and $k$. 
Finally, we find that we can match the critical line of transition B in the KNN and CB models when we set $a=1$ and $f = \langle n_C \rangle \, k$, thus showing that these expressions provide a good approximation for the mean effective interaction range and time \cite{SI}.

%-------------------------------------------------------------- FIGURE 3

\begin{figure}[t]
    \centering
    \setlength{\abovecaptionskip}{-0.0cm}
    % \begin{tabular}{cc}
    % \begin{subfigure}[b]{0.5\columnwidth}
      \includegraphics[width=0.49\columnwidth]{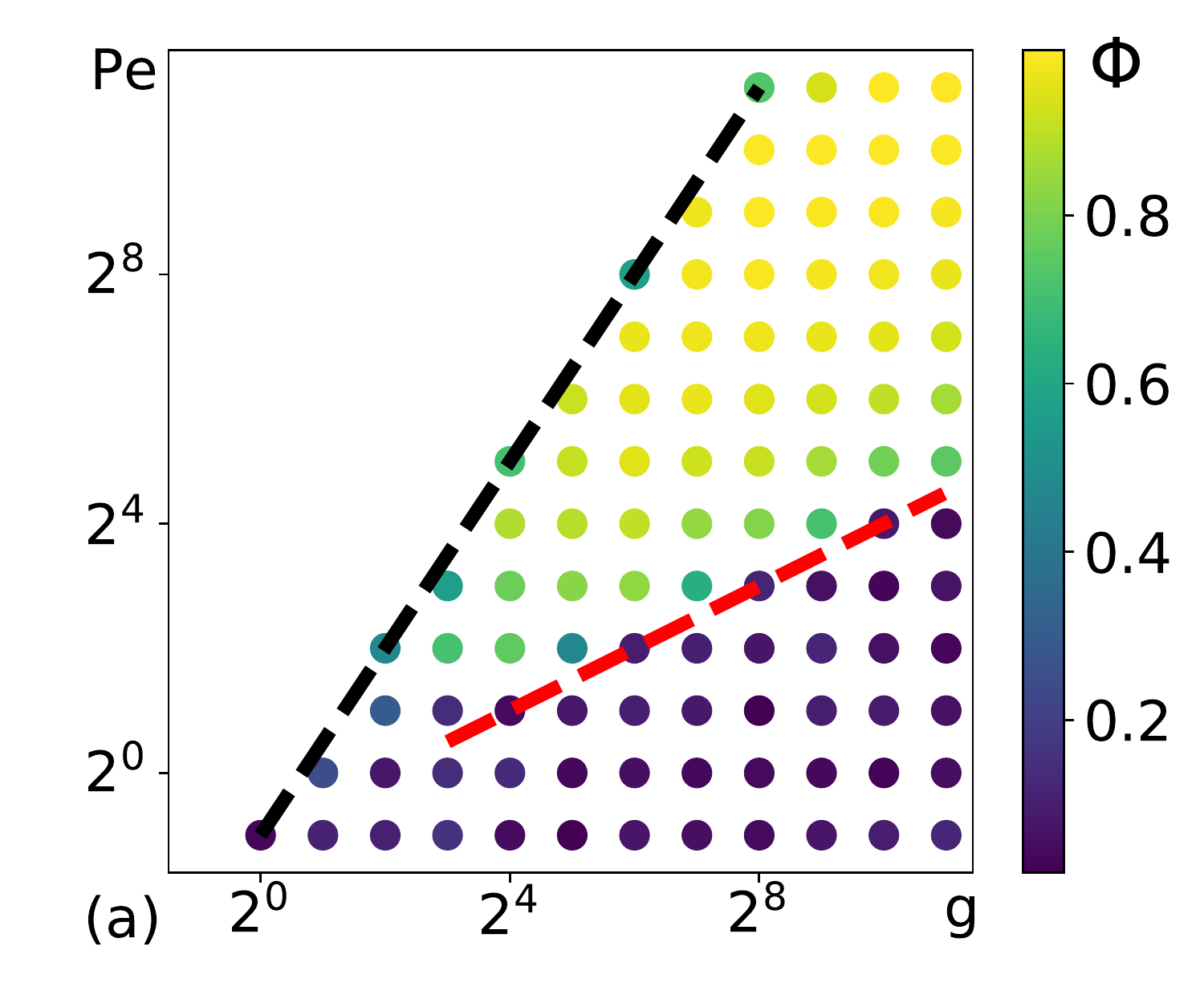}
    % \end{subfigure}
    % \begin{subfigure}[b]{0.5\columnwidth}
      \includegraphics[width=0.49\columnwidth]{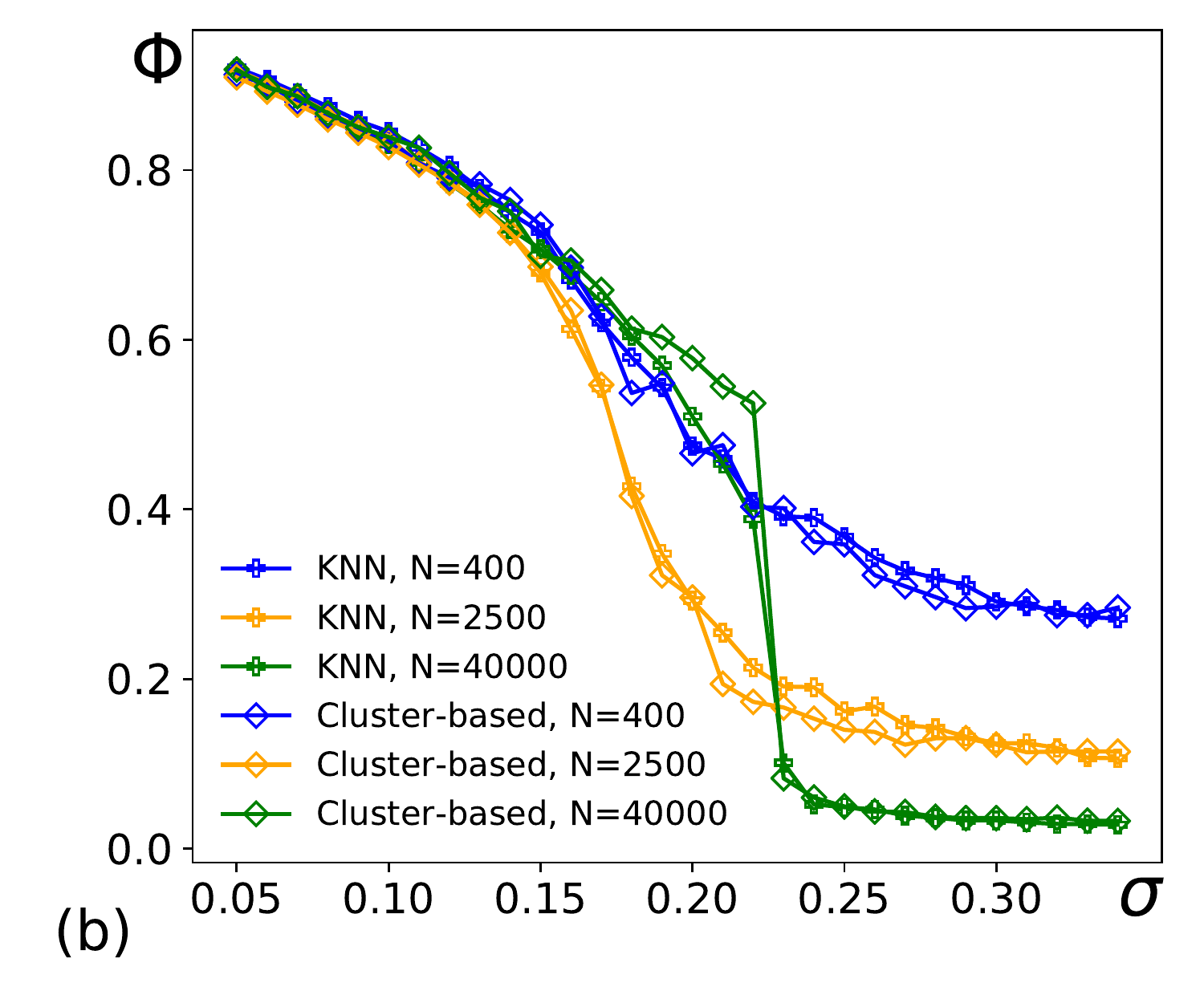}
    % \end{subfigure}

    % \end{tabular}
    \caption[Phase comparisons]{Comparison of the phases of the KNN model simulations (for $\rho=1$, $v_0=0.2$, $k=3$) and its corresponding coarse-grained CB model simulations.
    Both panels show that the KNN collective states and transitions are well captured by the CB model when using the parameter mapping detailed in the text.
    (a) Polar order phase diagram of CB simulations as a function of $\mathrm{Pe}$ and $g$, for $N=2500$. The overlaid dashed lines are the same as in Fig.~\ref{fig:diagrams}a, showing the clustering transition (black) and transition B (red).
    (b) Polar order as a function of noise $\sigma$ for KNN (fixing $\tau=0.3125$) and corresponding CB simulations of different sizes.
    Each $\Phi$ shows the mean polarization, averaged over $2\times10^4$ frames collected over $10$ time units for 10 independent runs after reaching the stationary state (except for the $N=40000$ case, for which we only used one simulation run).}
    \label{fig:order-disorder-transition-comparison}
\end{figure}

%-------------------------------------------------------------- FIGURE 3

Using the effective parameters derived above, we simulated the CB model to try to reproduce the phase diagram in Fig.~\ref{fig:diagrams}(a) with our coarse-grained system.
Figure \ref{fig:order-disorder-transition-comparison}(a) presents these results, where the overlaid dashed lines mark the same clustering line and transition B line displayed in Fig.~\ref{fig:diagrams}(a).
No simulation can be performed above the clustering line given by $\mathrm{Pe}/g^{1.5}=0.5$, since the CB model can only be applied when KNN clusters are formed.
The figure shows that there is a clear match between transition B in the KNN model and the standard topological order-disorder transition in the CB model.

%
%In fact, the CB model works best when $\mathrm{Pe}/g^{1.5} < 0.1$, in which case the cluster size does not depend on the system size (see Supplementary Information \ref{}).
%
In order to further compare the KNN and CB phases, we plot in Fig.~\ref{fig:order-disorder-transition-comparison}b the polarization $\Phi$ as a function of noise strength $\sigma$ for both models.
(Note that multiple parameters must be varied, according to the relationships above, to match the CB model to the corresponding KNN simulations as a function of $\sigma$.)
We find an excellent agreement between the curves, with only small deviations for the largest systems ($N=40000$). These can be attributed to the formation of high-density bands that we observe in the KNN simulations near the transition, which increase cluster heterogeneity and thus limit the validity of the CB description.
%

% By comparing the states in elongated space and square space (so that for same parameters band appears near transition B in the former but not appears in the latter), we confirm that clusters out of a band have significantly larger spatial size than clusters inside a band. The increasing heterogeneity violates the assumption of the cluster-based model that clusters behave uniformly.
% However, we note that the band structure does not significantly influence the distribution of cluster size, neither changes the order of the phase transition.

% Although we obtain $a$ and $f$ by fitting transition B, we note that it is not trivial that $a$ and $f$ have such simple forms and that the cluster-based model matches kNN model so well in non-band states with different system sizes.

%Our result gives a good example of individuals self-organizing into structures and influencing the macroscopic collective motion. 

% AQUI AQUI AQUI

%DISCUSSION PART (Pawel thinks it still needs revising)
Taken together, the results in Figs.~\ref{fig:nC-transition-1} and
\ref{fig:order-disorder-transition-comparison} show that KNN models with alignment interactions will form persistent clusters at a critical value of the control parameter $\mathrm{Pe}/g^{\alpha}$.
These clusters then behave as effective particles with metric interactions, which eventually start moving in different directions as $g$ is increased and Pe is reduced, thus losing global order.
We note that we observed a similar process when considering other alignment functions \cite{SI}.
This suggests that such scenario may be a generic feature of all models with KNN interactions, and thus amenable to an analytical description. 
For instance, a Smoluchowski-type equation could be used to derive the $\alpha$ exponent from first principles, as it was done to find the cluster size distribution in metric-based models \cite{peruani2010cluster}. 
However, new aggregation/disaggregation equations would be required in the KNN case, since we already know that the low noise limit behaves differently. Indeed, while large clusters are formed as $\sigma$ approaches zero in metric models, the opposite occurs in the KNN model.

In sum, we studied an alignment-based self-propelled particle model with KNN interactions in the ($g$,Pe) phase space and found two distinct types of order-disorder transition. 
The first one is continuous, occurs at low, fixed $g$ value, and is well captured by a standard mean-field description.
The second one, by contrast, is discontinuous and is governed by a single non-dimensional parameter $\mathrm{Pe}/g^{\alpha}$, with $\alpha\approx 1.5$. It is linked to the formation of dense, persistent clusters that behave as individual effective particles with metric interactions, and its properties thus resemble those of metric model transitions.

Our work shows how to reconcile conflicting reports claiming the presence of either a continuous \cite{chou2012kinetic} or a discontinuous \cite{martin2021fluctuation,rahmani2021topological} phase transition in the same type of metric-free system.
We find, in particular, that the parameters used in \cite{martin2021fluctuation} appear to match our transition B, thus explaining the observed discontinuity.
Finally, our results demonstrate a mechanism through which metric-like dynamics emerge from metric-free interactions (complementary to that in \cite{martin2021fluctuation,rahmani2021topological}), showing that characteristic length scales can still emerge in metric-free systems.
%

% \section*{Acknowledgement}
% Acknowledgement
% \vspace{0.3cm}

YZ is grateful for the financial support from the China Scholarship Council (CSC).
PR and YZ acknowledge funding by the Deutsche Forschungsgemeinschaft (DFG, German Research Foundation) through the Emmy Noether Programm - RO 4766/2-1 and under Germany’s Excellence Strategy – EXC 2002/1 “Science of Intelligence” – project number 390523135. 
The work of CH was partially funded by CHuepe Labs Inc and by Grant 62213 from the John Templeton Foundation.

\bibliographystyle{unsrt}
\bibliography{references}

\begin{thebibliography}{10}
\bibitem{endnote1}
Given the random nature of the simulations and the slow convergence rate in the
  low $\mathrm{Pe}$ regime, finding an exact $\alpha$ for all $\langle n_c
  \rangle$ values would demand very large numerical simulation that are beyond
  the scope of this work; we thus consider $\alpha=1.5$ to be a reasonable
  approximation of the exponent, as it is the closest simple rational number.
\bibitem{endnote2}
This ansatz is based on simplifying the problem by fixing the network structure
  in each cluster during collisions; the cluster alignment relaxation time
  $\tilde{\tau}$ must then be proportional to the individual particle
  relaxation time $\tau$; the ratio depends on $k$ and the cluster's inner
  structure; the average of the latter is determined by $\langle n_c \rangle$
  and $\langle r_c\rangle$, and can hence be expressed in terms of $\langle n_c
  \rangle$ only.
\end{thebibliography}

\end{document}